\documentclass[12pt]{iopart}
\usepackage{graphicx}
\usepackage[percent]{overpic}

\begin{document}

\title[]{Effect of fast electrons on the gain of a direct-drive laser fusion target}

\author{S~Yu~Gus'kov$^1$, P~A~Kuchugov$^{1,2}$\footnote{Present address: Keldysh Institute of Applied Mathematics of RAS, Miusskaya Sq.~4, Moscow, 125047, Russia}, R~A~Yakhin$^1$ and N~V~Zmitrenko$^2$}

\address{$^1$ P.N.~Lebedev Physical Institute of RAS, Leninskiy Av., 53, Moscow, 119991, Russia}
\address{$^2$ Keldysh Institute of Applied Mathematics of RAS, Miusskaya Sq., 4, Moscow, 125047, Russia}
\ead{pkuchugov@gmail.com}
\vspace{10pt}

\begin{abstract}
The results of numerical and theoretical studies of the gain of direct-drive inertial confinement fusion (ICF) target, which includes a kinetic description of energy transfer by laser-accelerated fast electrons, are presented. The range of initial temperature of fast electrons and fraction of laser energy contained in these particles were chosen based on the results of recent experiments at the National Ignition Facility (NIF).The effect of `wandering' of fast electrons is taken into account. It is due to a remoteness of the region of fast electron generation from the ablation surface of imploded target. As a result a significant fraction of particles do not fall into the compressed part of target. The `wandering' effect leads to decreasing the negative effect of fast electron generation on the target's gain.
\end{abstract}

%
\vspace{2pc}
\noindent{\it Keywords}: inertial confinement fusion, fast electrons, direct-drive targets, implosion, energy transfer.
%
%
%
%

\maketitle

\section{Introduction\label{sec:001}}

The effect of energy transfer by laser-accelerated fast electrons on target's compression and, as a result, on the thermonuclear gain occupies an important place in the problem of spark ignition of direct-drive ICF target. The most important characteristics of a source of fast electrons which are generated in extended plasma of  the evaporated part of a target (corona), are their total energy and particle’ temperature. These characteristics were investigated in recent experiments~\cite{Rosenberg2018} at NIF when a laser-plasma interaction was organized as close to implosion of the direct-drive target, intended for ignition. According to the results of~\cite{Rosenberg2018}, the generation of fast electrons is mainly due to Stimulated Raman Scattering (SRS) of  laser light and occurs near the plasma region with a quarter-critical density. About 1\% of laser energy is converted into the energy of fast electrons. Bearing in mind that 6-8\% of the laser energy is transformed into kinetic energy of the compressed part of direct-drive target, this value should be recognized as significant. The characteristic temperature of fast electrons was determined in~\cite{Rosenberg2018} to be approximately 50~keV. The mean free path of electrons with such an energy in a low-density corona far exceeds the corona's size. Thus, fast electrons have the ability to deliver the energy to the compressed part of target. Keeping in mind that they carry a significant fraction of laser energy, fast electron preheating can lead to a significant decrease in compression and, as a result, in thermonuclear gain of target.

The degree of gain degradation depends on how much of fast electron energy is deposited into the compressed part of target. The space-time distribution of fast electron energy, which is transmitted to a spark ignited direct-drive target during its implosion, was studied theoretically in~\cite{Guskov2019}. The calculations were performed with taking into account the time evolution of a thermodynamic state and hydrodynamic motion of imploded target. The ranges of fraction of laser energy contained in fast electrons, $5 \cdot 10^{-3} \leq \eta \leq 1.5 \cdot 10^{-2}$, and the temperature of fast electrons, $30~\mbox{keV} \leq T_h \leq 70~\mbox{keV}$, were chosen in accordance with the data of~\cite{Rosenberg2018}. It was assumed that fast electrons are generated in the region of corona with a density close to the quarter-critical density, with an isotropic angle distribution. The justification of an isotropy of fast electron source is the strong effect of laser light refraction in the region of particle generation. The main conclusion of~\cite{Guskov2019} is that the character of energy transfer by fast electrons to the compressed part of  target is strongly influenced by the effect, called in~\cite{Guskov2019} as the effect of  fast electron `wandering'. It is due to a remoteness of the region of fast electron generation from the ablation surface of imploded target and continuous reduction of this surface square. As a result a significant fraction of particles, moving in corona and repeatedly crossing it due to reflection in a self-consistent electric field, does not hit into the compressed part of the target. For a megajoule level target, the implosion velocity of which reaches 300-400~km/s, this effect leads to the fact that the fraction of fast electrons that can transfer their energy to the compressed part of target (the fraction of `heating' fast electrons) is small. For the typical direct-drive target~\cite{Belkov2015}, consisting of a shell with a layer of polystyrene ablator and a layer of frozen DT-ice, corresponding to an absorbed laser energy of 1.5~MJ, this fraction is only 12\% of the total number of fast electrons. As a result, the degree of negative influence of fast electron generation on the compression of a direct-drive ICF target is significantly reduced. So, when 1\% of laser energy is converted into the energy of fast electrons with energy of 50~keV, the final density of DT-fuel decreases compared to the case of the absence of generation of these particles by 2.4 times with taking into account the `wandering' effect and by 3.1 times without taking into account this effect.

This paper is devoted to a numerical and theoretical investigations of the effect of energy transfer by fast electrons on the gain of a direct-drive ICF target, designed for ignition. The gain is assumed as the ratio of the energy released in the fusion reactions to the absorbed energy of laser pulse. The numerical calculations were carried out using a one-dimensional hydrodynamic code, supplemented by the unit for calculating the energy source caused by the energy transfer by fast electrons of the Maxwell spectrum. The energy transfer to the target from fast electrons is calculated taking into account the `wandering' effect described above.

\section{Statement of the problem\label{sec:002}}

\begin{figure}[!ht]
  \centering
  \includegraphics[width=0.6\textwidth]{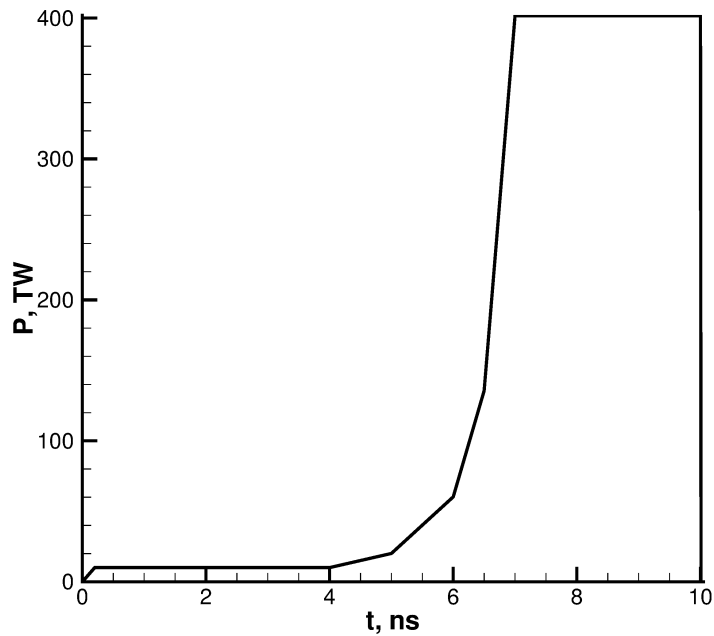}
  \caption{The dependence of the power of the laser pulse related to the absorbed energy, the fraction of which is 0.57. The image reused with permission from~\cite{Guskov2019}.}\label{fig:001}
\end{figure}

The studies were performed for the same direct-drive target, which was used in~\cite{Guskov2019} to study the features of energy transfer by fast electrons. The target corresponds to the irradiation conditions at the megajoule laser facility of the Russian laser project~\cite{Garanin2011}, which implies the action of a shaped laser pulse with the energy of 2.6~MJ of the 2$^{\mbox{nd}}$ harmonic of Nd-laser irradiation. The target in the form of a shell containing a layer of polystyrene ablator with a DT-ice layer frozen on its inner surface has external radius $R = 1597$~$\mu$m, thicknesses of CH-ablator layer and DT-ice layer are $\Delta_a = 34$~$\mu$m and $\Delta_{\mbox{DT}} = 149$~$\mu$m, respectively, the masses of the layers are $M_a = 1.12$~mg and $M_{\mbox{DT}} = 1.06$~mg. Inside the shell is a DT-gas with a density of $10^{-3}$~g/cm$^3$. Figure~\ref{fig:001} shows the time dependence of the pulse power for the absorbed laser flux with total energy of about 1.5~MJ. The pulse with a duration of 10 ns and a power contrast of 40 over a period of time of 0-4~ns has a power of 10~TW, which increases from 4 to 7~ns to a value of 400~TW, after which it does not change until the pulse termination.

The fraction of absorbed laser energy is 0.57. It was calculated using the Lagrangian RAPID code~\cite{Belkov2015,Rozanov1985}, which provides a joint solution of one-dimensional hydrodynamic and Maxwell equations, taking into account the inverse bremsstrahlung and resonant absorption mechanisms. Numerical  algorithm of laser radiation absorption in the RAPID code is based on the combination of ray and wave descriptions. According to this, a ray trajectory taking into account the refraction is constructed in the target corona and the Maxwell equations are solved in the vicinity (with double-wavelength thickness) near the turning point for the oblique incidence of a wave onto planar plasma with allowance for the $s$- and $p$-polarized wave components. The numerical simulations for considered target have been performed under the conditions of project~\cite{Garanin2011} with Gaussian laser beam with an aperture of 40~cm at a focal length of 660~cm. The ratio of the beam radius to the initial radius of target is equal to 1. In the RAPID-code calculation an irregular numerical mesh with mass intervals varying in a geometric progression is used. The mesh in CH-ablator was thickening to the outer boundary of target, in DT-ice layer it was thickening to the border of DT-gas. In the calculations of the considering target the total number of cells was 405. Of these, 40 cells were belonged to DT-gas, 198 -- to DT-ice layer, and 167 -- to ablator. At the initial time, the size of the outer cell in the ablator was 0.02~$\mu$m. By the end of the laser pulse, the number of cells in the plasma region with a density below the critical plasma density was 143. The time steps were regulated by the Courant condition. Their number for the duration of laser pulse amounted to about 12000.

The gain of the considered target, without taking into account its heating by fast electrons, is 21~\cite{Belkov2015}. It was calculated using the one-dimensional Lagrangian DIANA code~\cite{Zmitrenko1983}. The DIANA code ensures the solution in spherical geometry of equations of two-temperature hydro-dynamics with electron heat conduction, ion viscosity, absorption of laser radiation, bulk energy loss due to the radiation emission, fusion energy production using the kinetic description of the energy transfer by $\alpha$-particles and the real equation of matter state. Electron conductivity flux is calculated with a limiter factor equal to 0.06. The fraction of absorbed laser energy, computed by RAPID code, and power of incident laser pulse are used to calculate laser energy deposition to the target according to inverse bremsstrahlung mechanism. The fraction of thermal radiation in the energy balance does not exceed 2\% of the absorbed laser energy. The kinetic description of $\alpha$-particle transport in the DIANA code is based on the two-group angle approximation, called the `forward--backward' or Schwarzschild approximation (see, for example,~\cite{Zeldovich2002}). The rationale for using this method for $\alpha$-particle transport calculation in a spherical homogeneous DT-plasma was given in~\cite{Guskov1975}. To describe the thermodynamic properties of a laser plasma, the following model of the equation of state was used~\cite{Karpov1982,Karpov1988}. The ion component is described as an ideal gas in the approximation of the mean ion. To determine the ionization degree, the kinetic equation for the ionization due to electron impact, photo-recombination and triple recombination is solved. The electron component is described by thermal (as for ions) and elastic (`cold') components. Semi empirical relations are used to describe the elastic component~\cite{Karpov1982,Karpov1988}. Numerical method of solving corresponding equations with sources is based on work~\cite{Samarskii1992} and consists in approximation of differential equations by difference analogs and solving nonlinear difference equation using iterative algorithms. For discrete space representation it is used the irregular mesh, which is similar to one used in RAPID.

Numerical simulations of this work were carried out according to the same DIANA code, supplemented by the unit which calculates a source in the energy equation due to the energy transfer from fast electrons to plasma electrons in Coulomb collisions with taking into account scattering on ions. The use of the 2nd harmonic of Nd-laser radiation in the direct drive approach corresponds to a lower energy efficiency of the target implosion, namely, to a lower absorption of laser radiation and to a lower hydrodynamic coupling efficiency compared to the use of the 3rd harmonic radiation of Nd-laser radiation at NIF~\cite{Moses2011} and LMJ~\cite{Andre2000} facilities. In this case, the fraction  of absorbed energy is approximately 10-15\%, and the fraction of absorbed energy transferred to the kinetic energy of the shell is 5-10\% lower than in the case of the 3rd harmonic. This circumstance is balanced by the fact that the efficiency of conversion into the 2nd harmonic radiation is 30-40\% higher than into the 3rd one. In this regard, the parameters of the target under consideration and the characteristics of its implosion are close to the direct drive targets designed to use with 3rd harmonic radiation at NIF~\cite{Craxton2015} and LMJ~\cite{Brandon2014} facilities. However, the generation of fast electrons are significantly different for the cases of  2nd and 3rd harmonics.  At, approximately, the same intensities of radiation in the final parts of  laser pulse (about $1.2 \cdot 10^{15}$~W/cm$^2$) the coupling parameter $I \lambda^2$, with increasing of which the temperature of fast electrons increases, in the case of the 2nd harmonic radiation is 2.25 times larger than in the case  the 3rd harmonic radiation. In addition, the larger velocity of the target's corona expanding in the case of the 2nd harmonic radiation compared to the 3rd one can influence on the energy characteristics of fast electrons. However, this effect can be expected to be insignificant. Indeed, on the one hand, the growth in expanding velocity leads to a growth of spatial scale of the density gradient in the region of fast electron generation and, as a consequence, to a decrease in the threshold of  laser intensity  corresponding to the development of plasma instabilities. On the other hand, this leads to a greater removal of the region of fast electron generation from the surface of the compressed part of the target and to a increase, thereby, of the fraction of `wandering' fast electrons.

For the purpose of the most general formulation of the problem these features of fast electron generation are taken into account by choosing a sufficiently wide ranges of fast electron temperature and efficiency of laser energy conversion into fast electron energy. These ranges are chosen the same as in~\cite{Guskov2019}, which in turn were selected on the basis of the results of experiments~\cite{Rosenberg2018}. As in~\cite{Guskov2019}, following the results of~\cite{Rosenberg2018}, it is believed that fast electrons are generated in the corona region with a density close to the quarter-critical plasma density due to SRS. It is assumed that the spectrum of angular distribution of the generated particles is isotropic. The argument of this assumption is the large role of laser radiation refraction  in the extended corona of direct drive targets intended for ignition~\cite{Guskov2019}. However, in contrast to the paper~\cite{Guskov2019}, in which the fast electrons were monoenergetic, in this paper we consider the particles with Maxwellian spectrum. Thus, the considered ranges of fast electron temperature and fraction of laser energy contained in these particles are
\begin{equation}\label{eq:001}
  30~\mbox{keV} \leq T_h \leq 70~\mbox{keV},
\end{equation}
\begin{equation}\label{eq:002}
  5 \cdot 10^{-3} \leq \eta \leq 1.5 \cdot 10^{-2},
\end{equation}
where $\eta$ is defined as the ratio of the energy $E_h$ contained in fast electrons to the energy of laser radiation reaching a region with a density equal to a quarter of the critical density. The ranges are located on both sides of the values $T = 50$~keV and $\eta = 0.01$, which correspond to the data of the experiment~\cite{Rosenberg2018}, performed when a flat target is irradiated with a nanosecond pulse of the 3rd harmonic of the Nd laser radiation. Parts of these ranges with higher values of temperature and energy fraction $\eta$ in comparison with the experimental data~\cite{Rosenberg2018}, meet the possibility of exceeding the values recommended in~\cite{Rosenberg2018}, when using the 2nd harmonic radiation, and the parts with lower values $T$ and $\eta$ meet the possibility of reducing these values due to the fact that the length of the spherically exploded corona is shorter than in the flat case.

The generation of fast electrons takes place in the period of the high-intensity part of laser pulse, in the period of time $\tau_h = 3$~ns from the time moment of $t_0 = 7$~ns until the end of laser pulse, $\tau_L = 10$~ns. The laser energy corresponding to this period of time is 75\% of the total laser energy that is 2.1~MJ. Approximately 90\% of the laser energy reaches the surface with a quarter-critical density. The rest of energy is absorbed in the corona's region with density less than the quarter-critical density. Thus, the range of energy contained in fast electrons, in accordance with the range~(\ref{eq:002}) parameter $\eta$ variation is $9.45 \leq E_h~(\mbox{kJ}) \leq 28.3$.

\begin{figure}[!ht]
  \centering
  \includegraphics[width=0.6\textwidth]{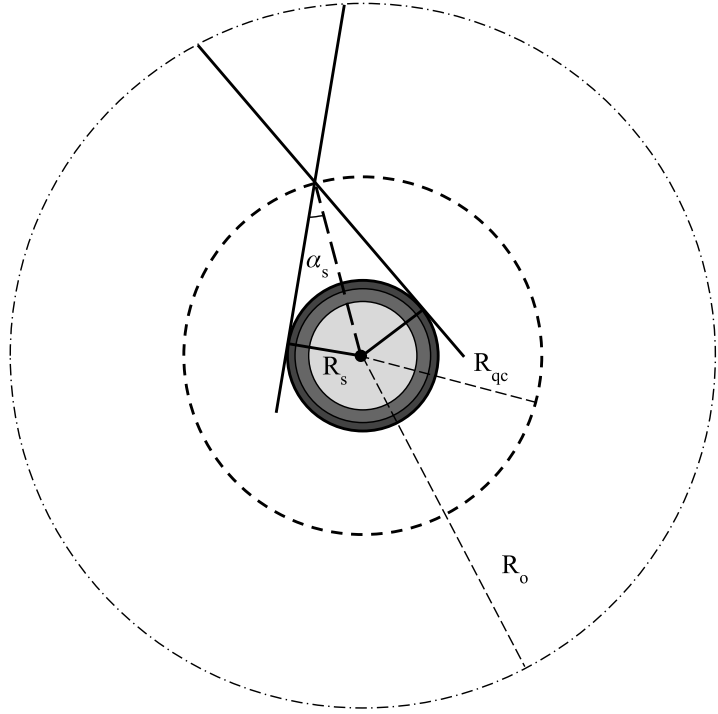}
  \caption{Scheme of imploded target regions related to birth and motion of fast electrons. The fast electron are generated  on the surface with a quarter-critical density (dashed circle). The two central spherical layers denote the non-evaporated part of the ablator and the DT-layer. The external dash-dotted circle denotes the surface of reflection of fast electrons inside the target due to the self-consistent electric field. The image reused with permission from~\cite{Guskov2019}.}\label{fig:002}
\end{figure}

The calculation of energy source due to fast electron heating in the period of $\tau_h = 3$~ns is based on the model described in~\cite{Guskov2019}, which takes into account the `wandering' effect. The fraction of fast electrons that heat the compressed part of target $\delta N_h(t)$ is calculated at each time step in the period of $t_0 \leq t \leq \tau_L$. These are particles that have the ability to get to the ablation surface. One half of these particles are particles moving after their generation to the center of target in the solid angle $\Delta \Omega_s$, whose vertex lies on a circle corresponding to the position of the region with a density equal to the quarter-critical plasma density and which rests on the sphere corresponding to the surface of the non-evaporated part of target (Fig.~\ref{fig:002}). The second, equal to the first (in the approximation of specular reflection, half of the `heating' fast electrons are the particles, which after their birth move in a solid angle, vertical to one described above. They are first decelerated in the corona with a density below the quarter-critical density and, after reflection in the self-consistent electrical field, fall on the border of the non-evaporated part of the target. The energy deposited from the `heating' fast electrons is calculated in accordance with the particle mass range and the areal density of  numerical cell. The rest, `wandering', fast electrons transfer their energy only to the corona, repeatedly crossing it as a result of reflection from its outer boundary. Fast electrons with a temperature in the range~(\ref{eq:001}) transfer their energy to the corona in 5-10 flights, depending on their initial energy. Taking into account the multi-span nature of energy transfer, the spatial distribution of energy deposited from `wandering' fast electrons is described in the approximation of uniform distribution over the corona's mass. The scattering of fast electrons due to the stochastic origin of the process does not change the ratio of the numbers of `heating' and `wandering' particles.

At a constant power of fast electron source $q_0$, the power of the source of `heating' fast electrons is given by~\cite{Guskov2019}
\begin{equation}\label{eq:003}
  q_h(t) = \delta_h(t) q_0,
\end{equation}
where
\begin{equation}\label{eq:004}
  \delta_h = \frac{2 \Delta \Omega_s}{4 \pi} = 1 - \cos{\alpha_s}.
\end{equation}
In turn, $\cos{\alpha_s} = \sqrt{1 - \left(R_s / R_{qc}\right)^2}$. The radii of the corresponding surfaces are determined and the value $\delta_h$ is calculated at each time step. The average implosion velocity of considered target is 350~km/s.  In this case, as it was shown in~\cite{Guskov2019}, the fraction of all `heating' fast electrons $\delta N_h$ is only 12\% of the total number of fast electrons $N_0$.

To determine the amount of energy transferred by fast electrons to the target, the model of stopping power of  electron plasma component together with scattering power of  ionic plasma component is used~\cite{Sivukhin1966,Ribeyre2013}. The expressions used for the fast electron mean free path length are close to the calculated results and the approximation formulas of~\cite{Deutsch1996,Li2004,Solodov2008,Atzeni2008,Atzeni2008a}. Fast electron mass range with taking into account the braking in collisions with free and bounded electrons as well as scattering in the collisions with ions is calculated as
\begin{equation}\label{eq:005}
  \mu = \frac{A m_p E_h^2 R}{4 \pi e^4 \gamma \left[Z_i \Lambda_{fe} + (Z - Z_i) \Lambda_{be}\right]}.
\end{equation}
Here, $E_h$ is fast electron energy, $\gamma = 1 + E_h / m_e c^2$ is relativistic factor, $m_e$ and $e$ are electron mass and charge, respectively; $c$ is light speed, $A$ is ion mass number, $Z_i$ and $Z$ are, respectively, the ion charge and the charge of a completely ionized ion; $m_p$ is the proton mass; the scattering factor reads
\begin{equation}\label{eq:006}
  R = \left[1 - \exp{\left(-\frac{\gamma \pi^2}{4 Z_i^2} \frac{Z_i \Lambda_{fe} + \left(Z - Z_i\right) \Lambda_{be}}{\Lambda_i}\right)}\right]^{1/2}.
\end{equation}
For the plasmas of different parts of considered target values of Coulomb logarithms changes in the intervals: $\Lambda_{fe} = 11-8$,  $\Lambda_{be} = 6-5$, $\Lambda_i = 8-3$.
At each time step $\Delta t_n$ the energy, deposited from $\delta N_h(t_n)$ fast electrons, is calculated by averaging over the Maxwell spectrum of the transmitted energy depending on the areal density of the cell and added as a source term into the energy equation of electron subsystem.

In order to establish the role of the `wandering' effect on the gain, two series of numerical calculations of the above described target were performed. One of them is taking into account the effect of `wandering', the second is without taking into account this effect. In the latter case all generated fast electrons are the `heating' ones. The results and their discussion are presented in the next section.

\section{Effect of fast electrons on target's compression and gain\label{sec:003}}

\begin{figure}[!ht]
    \centering
    \begin{minipage}[b]{0.45\textwidth}
        \begin{overpic}[width=\textwidth]{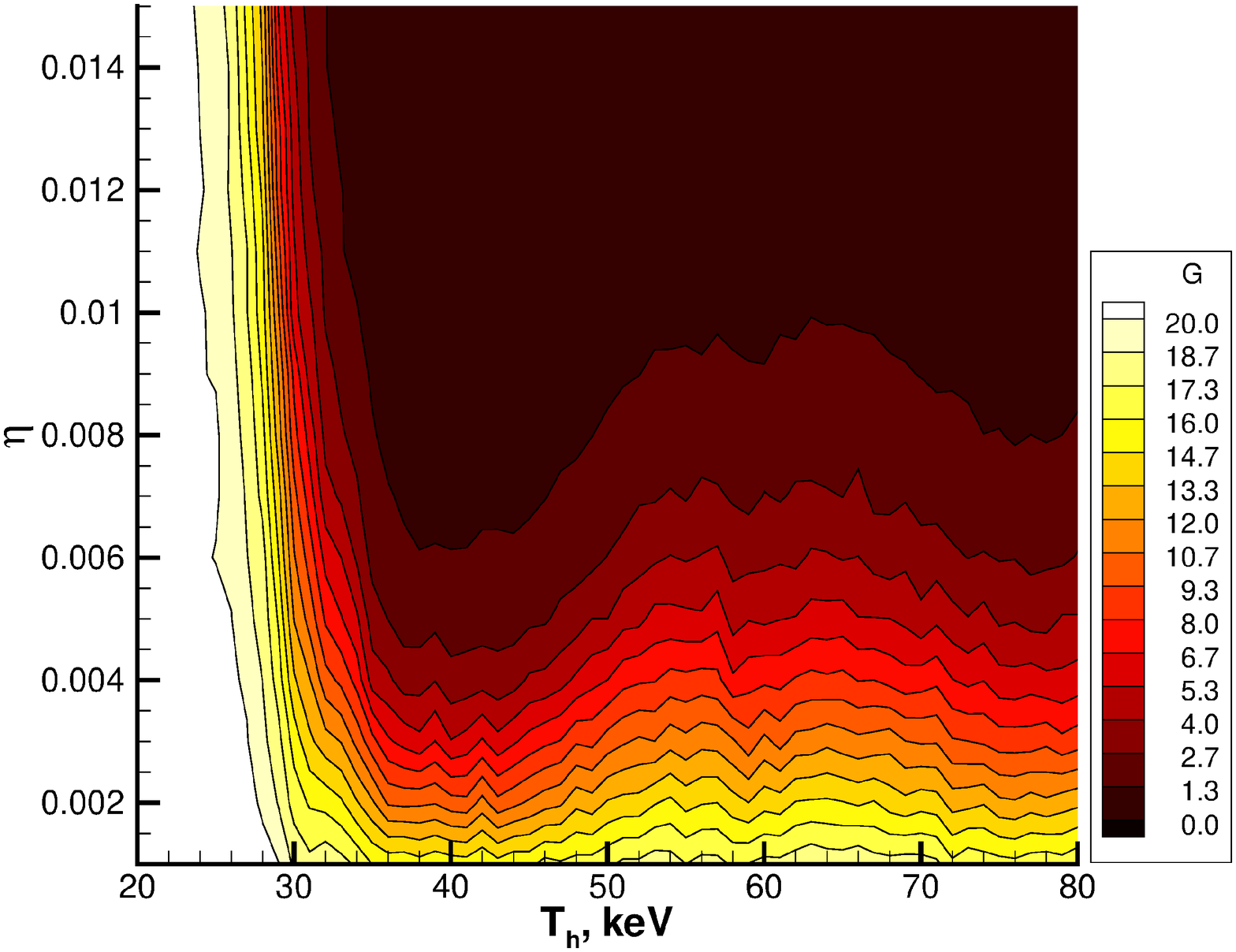}
            \put(93,72){(a)}
        \end{overpic}
    \end{minipage}
    \quad
    \begin{minipage}[b]{0.45\textwidth}
        \begin{overpic}[width=\textwidth]{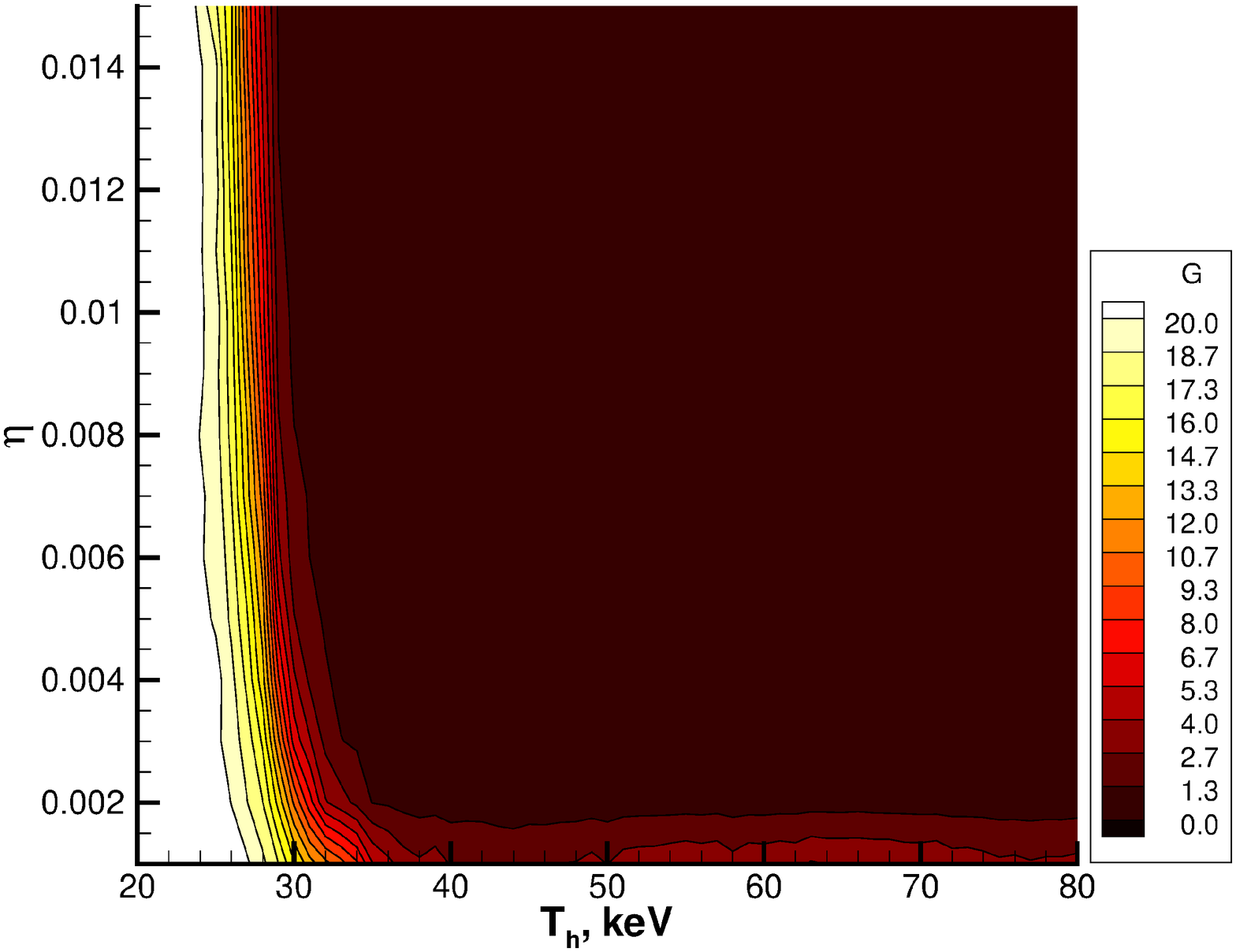}
            \put(93,72){(b)}
        \end{overpic}
    \end{minipage}
    \caption{Contour lines of the gain at various temperatures and laser energy conversion into  fast electron energy: (a) with taking into account the `wandering' effect; (b) without taking into account the `wandering' effect.}\label{fig:003}
\end{figure}

\begin{figure}[!ht]
    \centering
    \begin{minipage}[b]{0.45\textwidth}
        \begin{overpic}[width=\textwidth]{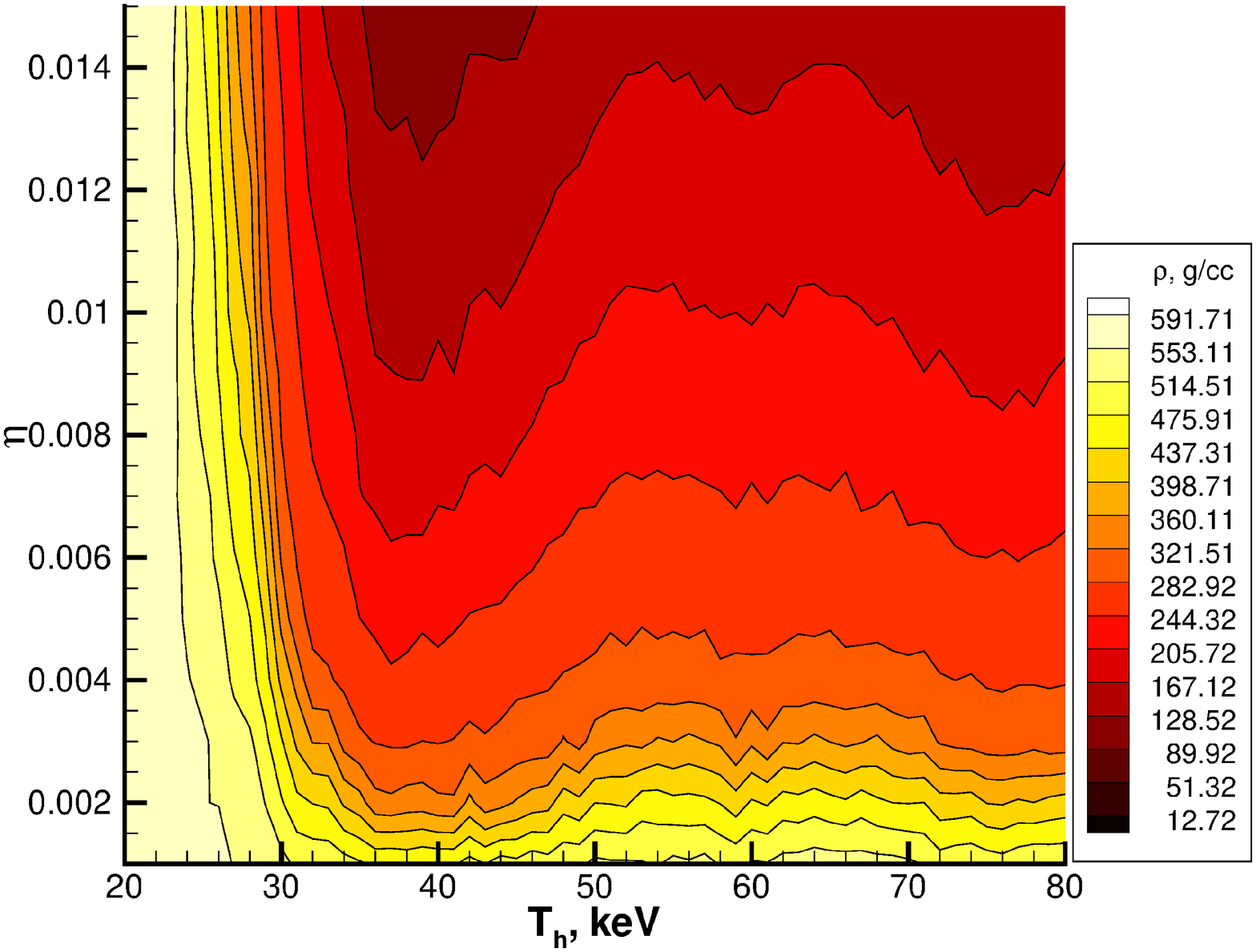}
            \put(93,72){(a)}
        \end{overpic}
    \end{minipage}
    \quad
    \begin{minipage}[b]{0.45\textwidth}
        \begin{overpic}[width=\textwidth]{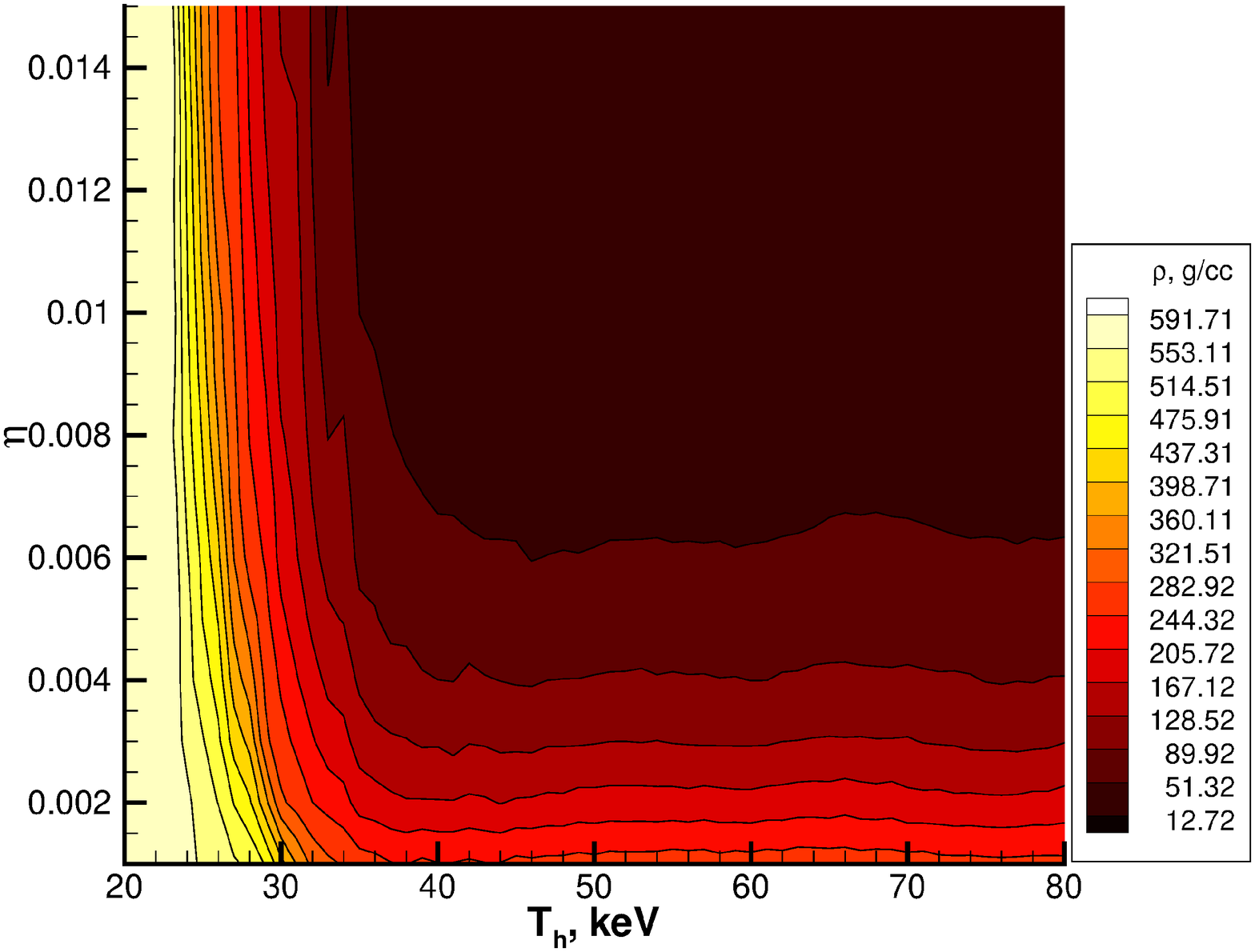}
            \put(93,72){(b)}
        \end{overpic}
    \end{minipage}
    \caption{Contour lines of the maximum density in DT-fuel at various temperatures and laser energy conversion into fast electron energy: (a) with taking into account the `wandering' effect; (b) without taking into account the `wandering' effect.}\label{fig:004}
\end{figure}

Figures~\ref{fig:003} and~\ref{fig:004} show the contour lines of the gain (Fig.~\ref{fig:003}) and maximum density in DT-fuel (Fig.~\ref{fig:004}) corresponding to  the results of numerical calculations made with and without the `wandering' effect at the different fast electron temperatures and laser energy conversion into fast electron energy, taken in accordance with the ranges~(\ref{eq:001}) and~(\ref{eq:002}).


First of all, it should be noted that up to the temperature of approximately $T_h \approx 28$~keV in the calculations, taking into account the `wandering' effect, and $T_h \approx 25$~keV in the calculations without the `wandering' effect, the negative effect of energy transfer by fast electrons on the gain is weak in whole range of efficiency of laser energy conversion into fast electron energy. At the temperatures below these values, fast electrons almost do not heat the compressed part of the target. They transfer energy to the evaporated part of target. The DT-fuel density and the efficiency of energy transfer to the kinetic energy of the compressed part of target remain almost the same as in the calculation without taking into account the fast electron generation which are, respectively, $\rho_{\max} \approx 700$~g/cm$^3$ and $\sigma \approx 6.67$\%.

\begin{figure}[!ht]
    \centering
    \begin{minipage}[b]{0.45\textwidth}
        \begin{overpic}[width=\textwidth]{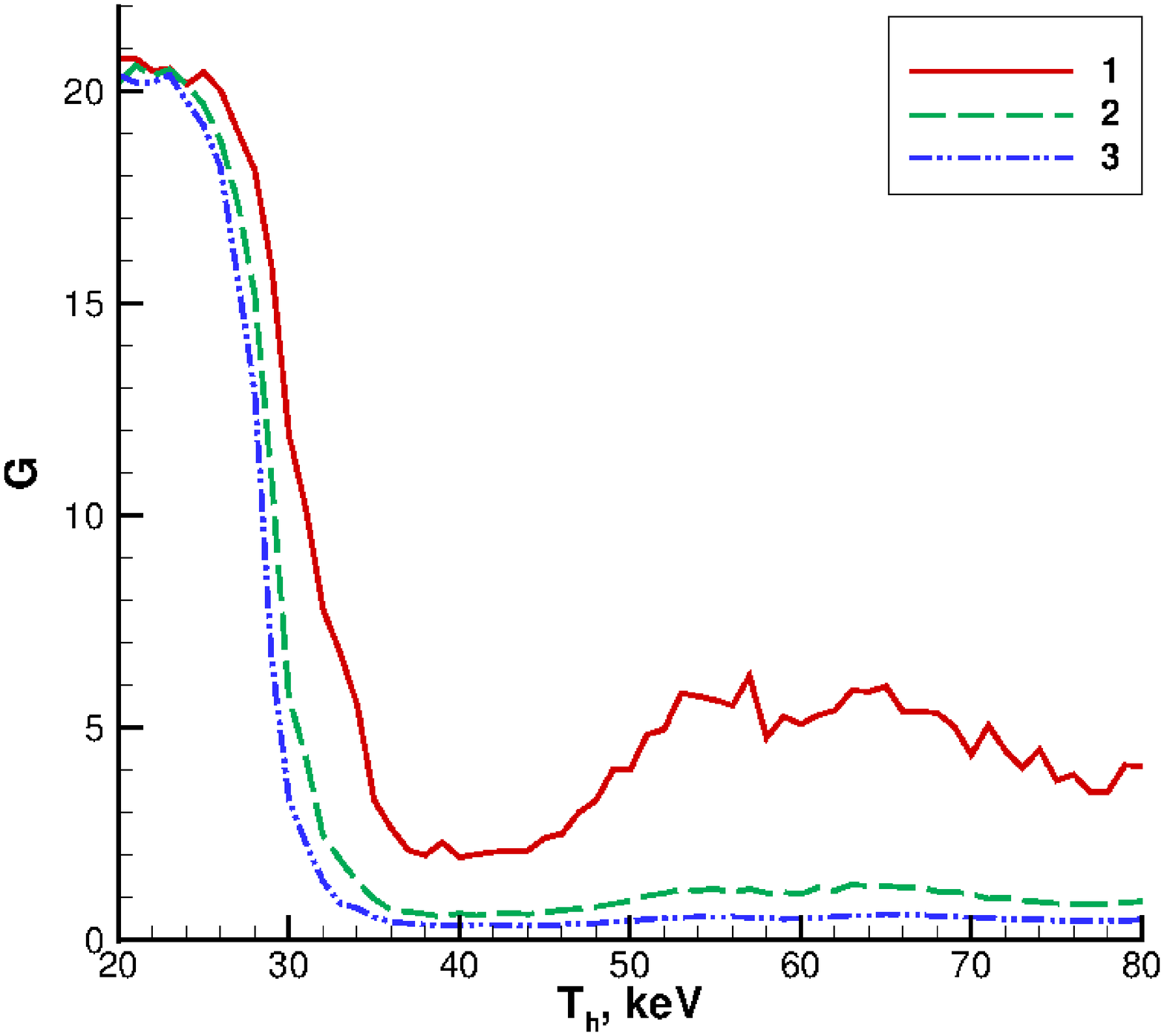}
            \put(47,83){(a)}
        \end{overpic}
    \end{minipage}
    \quad
    \begin{minipage}[b]{0.45\textwidth}
        \begin{overpic}[width=\textwidth]{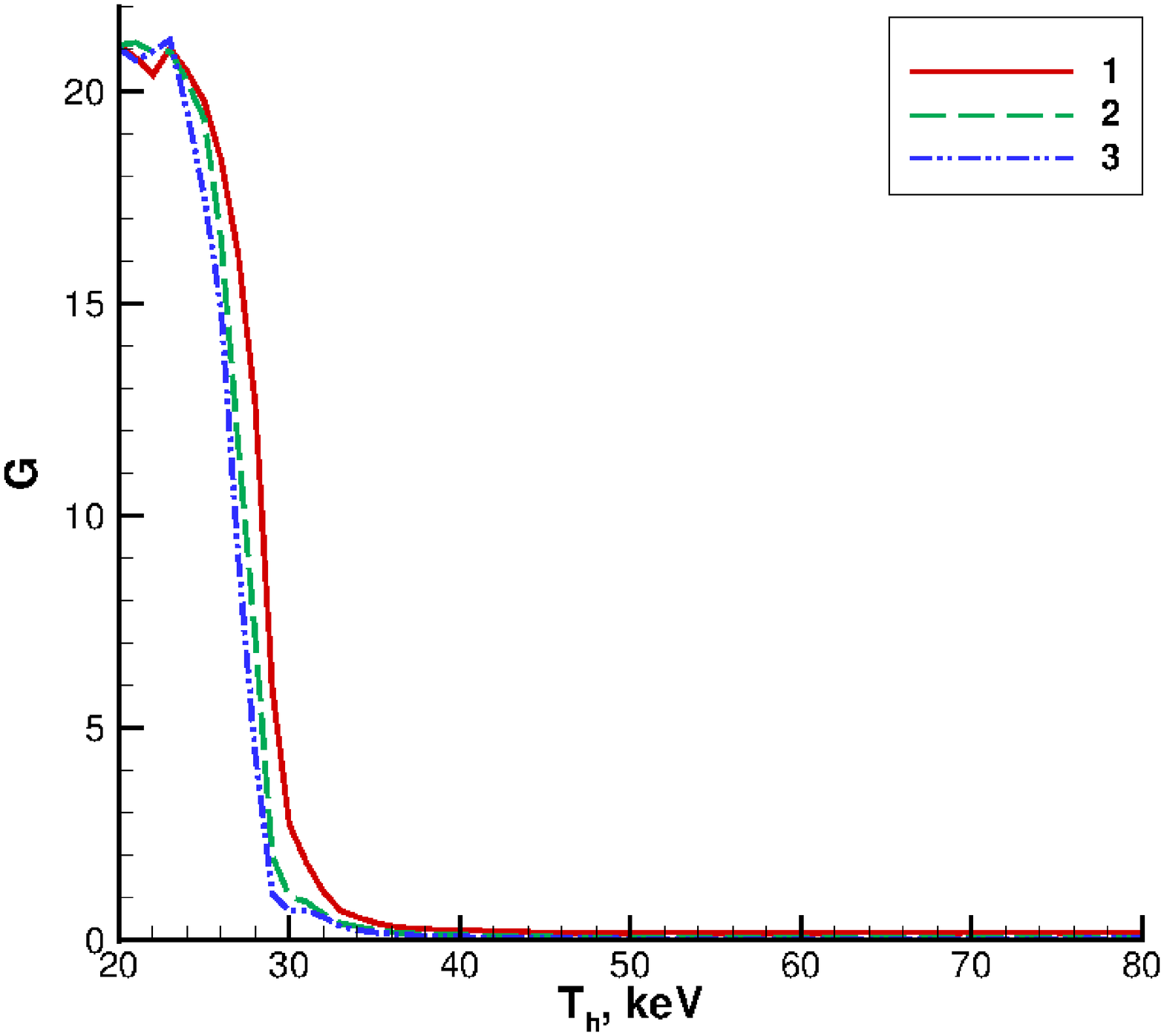}
            \put(47,83){(b)}
        \end{overpic}
    \end{minipage}
    \caption{Dependencies of the gain, taking into account the `wandering' effect (a) and without consideration of this effect (b) on the fast electron temperature $T_h$ for the different fractions of laser energy contained in fast electrons. Curves 1--3 correspond to $\eta$ values $0.005$, $0.01$ and $0.015$.}\label{fig:005}
\end{figure}

At the temperatures $T_h$ above 25-28~keV, fast electrons begin to heat the compressed part of target, including DT-fuel. Due to this, the negative effect of fast electrons begins to manifest itself very quickly  with increasing  the temperature $T_h$.
This effect is especially strong in the calculations without taking into account the `wandering' effect, when it is manifested even at $T_h = 34$~keV. The target does not ignite at any value of laser energy conversion into fast electron energy from the range~(\ref{eq:002}). The DT-fuel density drops to values of $100-150$~g/cm$^3$.The `wandering' effect significantly reduces the negative effect of fast electrons on the gain. However, there is a range of $T_h$ and $\eta$ values in which the target does not ignite, even if the `wandering' effect is taken into account. At $T_h = 34$~keV, the target does not ignite when $\eta > 0.08$. At $\eta = 0.05$, the target ignites, although the gain is about 5 that  is almost 4 times smaller than in the calculation without fast electron generation ($G = 21$). The non-ignition region can be defined as $\eta >0.005$ and $35 < T_h < 55$~keV. The nonmonotonic character of the gain and final DT-fuel density dependencies on fast electron temperature at a given fraction of laser energy contained in fast electrons $\eta$ attracts oneself attention. This character is particularly pronounced in the calculations taking into account the `wandering' effect. The dependencies of the gain on temperature $T_h$ in the calculations with and without the `wandering' effect are shown in Fig.~\ref{fig:005}. The gain, taking into account the `wandering' effect, has a minimum near the temperature of about $T_h \approx 40$~keV. Then it increases when the temperature grows to the value of about $T_h \approx 64$~keV with a subsequent gradual decrease. Moreover, the characteristic temperatures $T_h$ corresponding to the minimum and maximum of dependencies are almost the same for different values of energy transformation factor $\eta$. The latter circumstance points that the nature of such dependencies is determined by the ratio of fast electron mass range, which grows with temperature $T_h$, and areal density along different paths through the target.

\begin{figure}[!ht]
    \centering
    \begin{minipage}[b]{0.45\textwidth}
        \begin{overpic}[width=\textwidth]{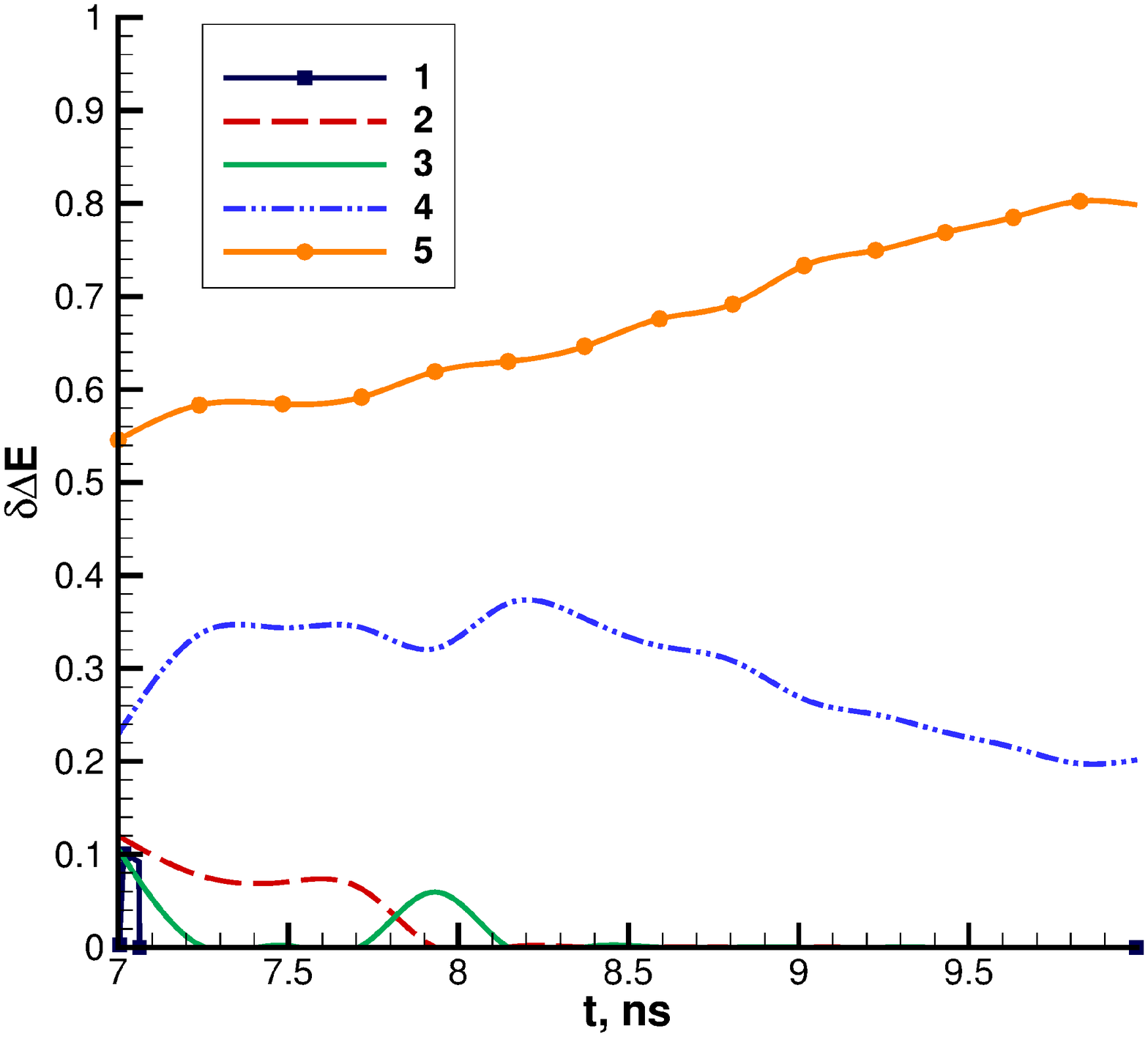}
            \put(92,83){(a)}
        \end{overpic}
    \end{minipage}
    \quad
    \begin{minipage}[b]{0.45\textwidth}
        \begin{overpic}[width=\textwidth]{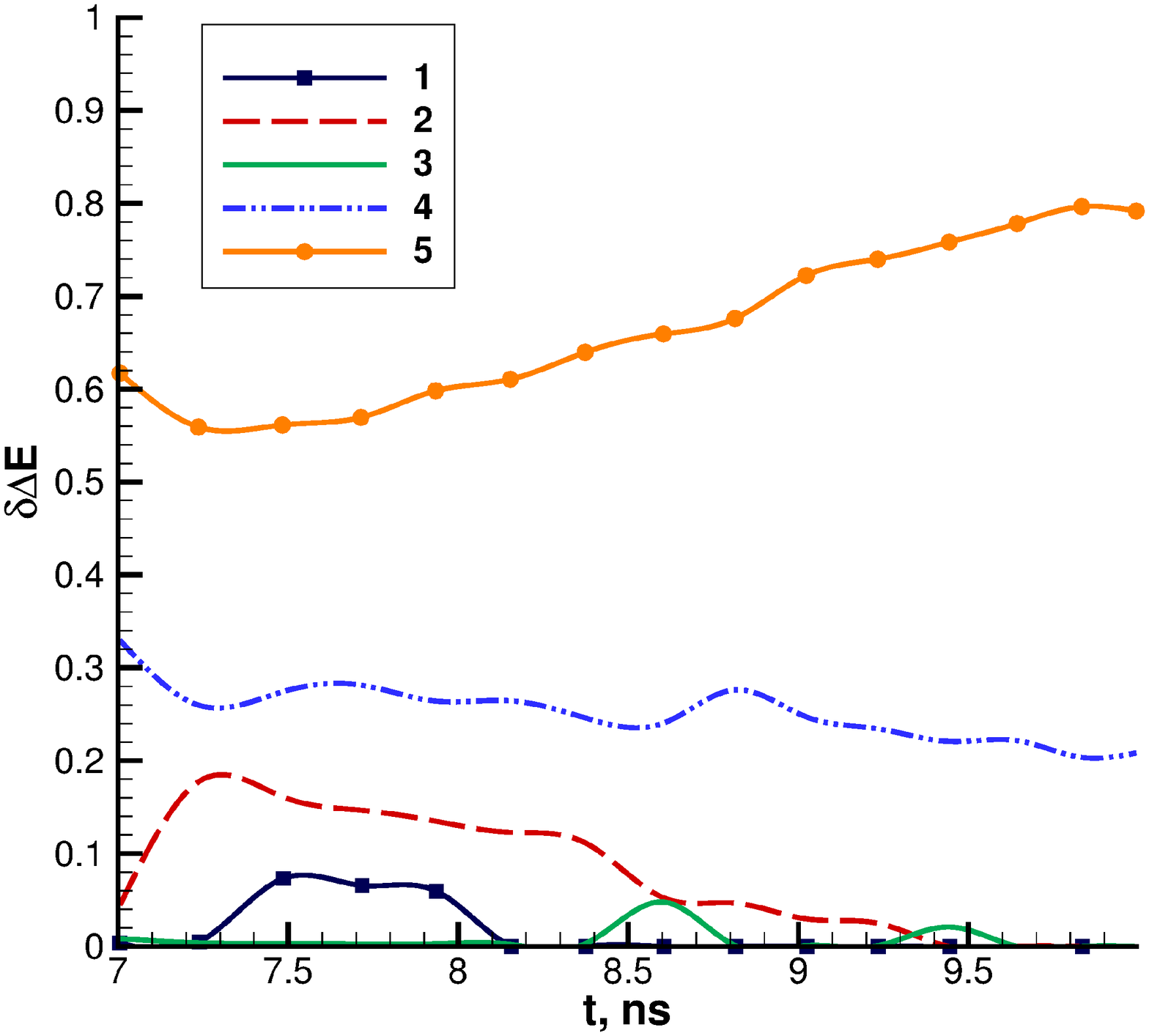}
            \put(92,83){(b)}
        \end{overpic}
    \end{minipage}
    \quad
    \begin{minipage}[b]{0.45\textwidth}
        \begin{overpic}[width=\textwidth]{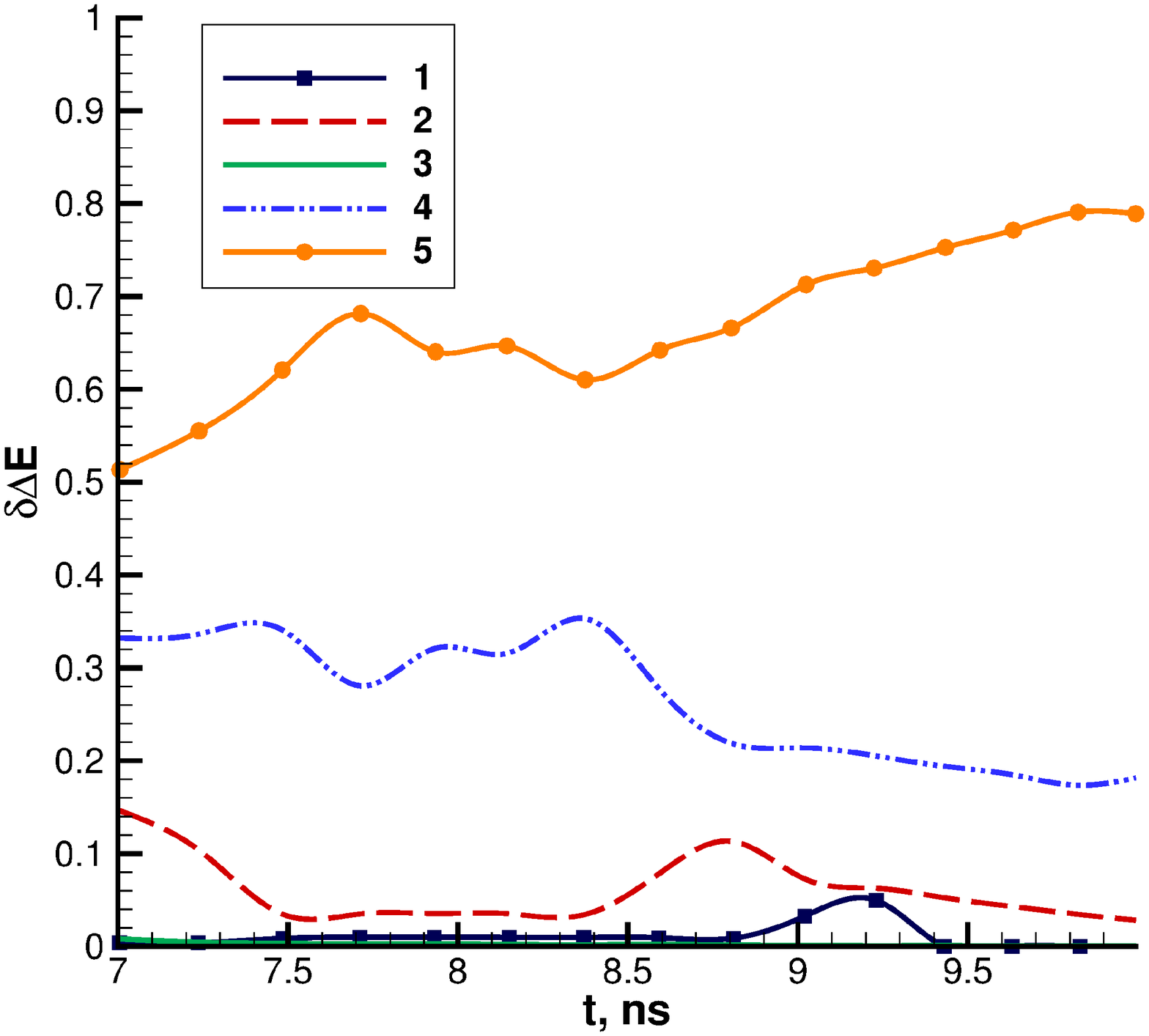}
            \put(92,83){(c)}
        \end{overpic}
    \end{minipage}
    \quad
    \begin{minipage}[b]{0.45\textwidth}
        \begin{overpic}[width=\textwidth]{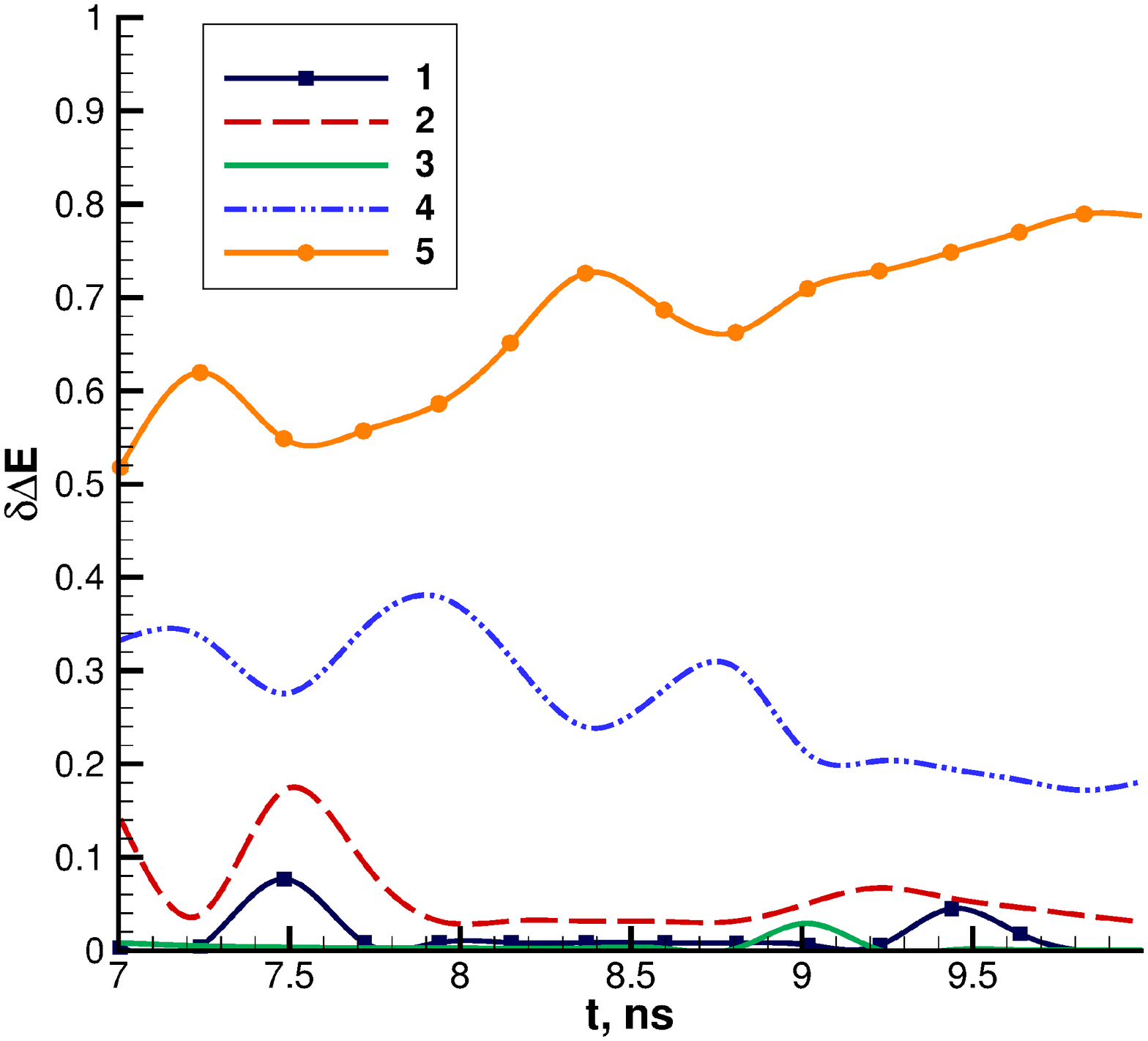}
            \put(92,83){(d)}
        \end{overpic}
    \end{minipage}
    \caption{The fractions of fast electron energy  transferred to the different parts of target. The temperature of fast electrons is (a) $T_h =30$~keV, (b) $T_h =40$~keV, (c) $T_h = 60$~keV and (d) $T_h = 70$~keV; the energy conversion  factor is $\eta = 0.01$. Curves 1--5 correspond, respectively, to inside portion of DT-fuel with a 10\% mass, to whole DT-fuel, to non-evaporated part of CH-ablator, to corona region from the evaporation boundary to the surface with a critical density and to corona region with a density less than the critical one.}\label{fig:006}
\end{figure}

It is confirmed with an additional analysis of the changing the temporal evolution of spatial distribution of the energy transferred from fast electrons in dependence of their temperature. Fig.~\ref{fig:006} shows the time dependencies of the fractions of fast electron energy transferred to the different parts of target, namely, to inside portion of DT-fuel with a 10\% mass which is, mainly, involved in the hot-spot formation, whole DT-fuel, non-evaporated part of CH-ablator, corona region from the evaporation boundary to the surface with critical density and corona region with a density less than critical one. These data correspond to the energy conversion factor  $\eta = 0.01$ for four characteristic values of the temperature $T_h$, namely, for $T_h = 30$~keV, $T_h = 40$~keV, corresponding to the minimum gain, $T_h = 60$~keV, corresponding to the local maximum gain, and $T_h = 70$~keV.

\begin{figure}[!ht]
  \centering
  \includegraphics[width=0.6\textwidth]{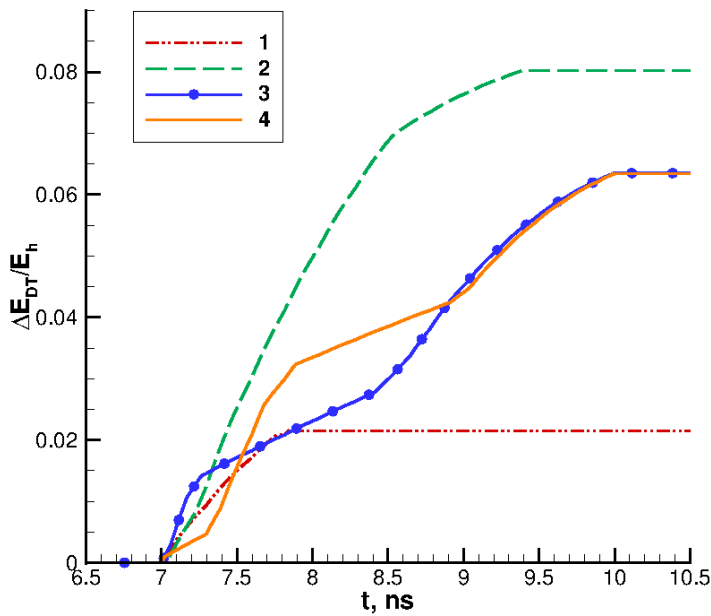}
  \caption{The dependencies of the energy transmitted to the DT-fuel up to the current time moment. Curves 1--4 correspond to fast electrons temperature $30$~keV, $40$~keV, $60$~keV and $70$~keV respectively. The fraction of laser energy converted to fast electrons is 1\%.}\label{fig:007}
\end{figure}

The initial decrease in the gain with an increase in the temperature $T_h$ corresponds to an increase in the time during which fast electrons are able to transfer their energy to the DT-layer. A further increase in fast electron temperature is accompanied by a decrease in the absolute value of the energy transferred to the DT-layer. It is due to the fact that the maximum of the fraction of deposited energy shifts to later times, corresponding to a significant decrease in the power of the `heating' fast electrons. This continues until the temperature of the fast electrons becomes sufficient to reach the DT-fuel on the subsequent passage through the target. This leads to an increase in the fraction of energy transferred to the DT-layer at the initial time moments and, accordingly, to a decrease in the gain. It should be noted that energy transfer from fast electrons to inside portion of DT-fuel, mainly, involved in the hot-spot formation (curves 1 in Fig.~\ref{fig:006}) for the cases of $T_h = 30$~keV and 40~keV occurs at initial time period, which is much smaller than the period of deposition to whole DT-fuel. With increasing temperature this period extends over the entire period of action of the high-intensity part of the pulse. Fig.~\ref{fig:007} shows the time dependencies of the absolute energy transferred to the DT-layer for the four cases considered above.

The fractions of energy transferred to the different parts of target are determined by the ratio of fast electron mass range and areal density of the corresponding region. This can be illustrated by the example of calculation with $T_h = 40$~keV and $\eta = 0.01$. A time moment of 7.5~ns is considered. According to the numerical data, the areal densities are of the order of: $5 \cdot 10^{-3}$~g/cm$^2$ in the DT-layer, $2 \cdot 10^{-3}$~g/cm$^2$ in the compressed CH-layer, $1.5 \cdot 10^{-3}$~g/cm$^2$ in the evaporated CH-ablator of supercritical density. In these regions, the mass ranges of a fast electron with an initial energy of 60~keV, equal to the root-mean-square energy of the particles of the Maxwell spectrum in the case under consideration, are, respectively, $5 \cdot 10^{-3}$~g/cm$^2$, $1.6 \cdot 10^{-2}$~g/cm$^2$ and $3 \cdot 10^{-3}$~g/cm$^2$. All generated fast electrons are moving in the evaporated ablator. So the energy fraction, which is deposited in this part of the target, can be simply calculated according to~\cite{Guskov2019} as $1 - \sqrt{1 - \rho r / \mu} \approx 0.3$. This value coincides with fairly good accuracy with the calculated value given in Fig.~\ref{fig:006}b (curve 3). In the compressed CH-ablator, fast electrons almost don't leave their energy due to its small areal density compared with the mass range in this region. The fraction of `heating' fast electrons at the  time of 7.5~ns is 0.22, which can be obtained using expressions~(\ref{eq:003}) and (\ref{eq:004}). Only this small fraction of fast electrons is passed into DT-fuel. These electrons have energy smaller than initial and leave the rest of their energy in this part of the target because of close values of areal density and mass range. The fraction of energy, which is deposited into DT-fuel, can be calculated as a product of fraction of `heating' fast electrons and the fraction of energy after passing the ablator and is about 0.15. This values is also close to the calculated one(curve 1 in Fig.~\ref{fig:006}b).

\begin{figure}[!ht]
    \centering
    \begin{minipage}[b]{0.45\textwidth}
        \begin{overpic}[width=\textwidth]{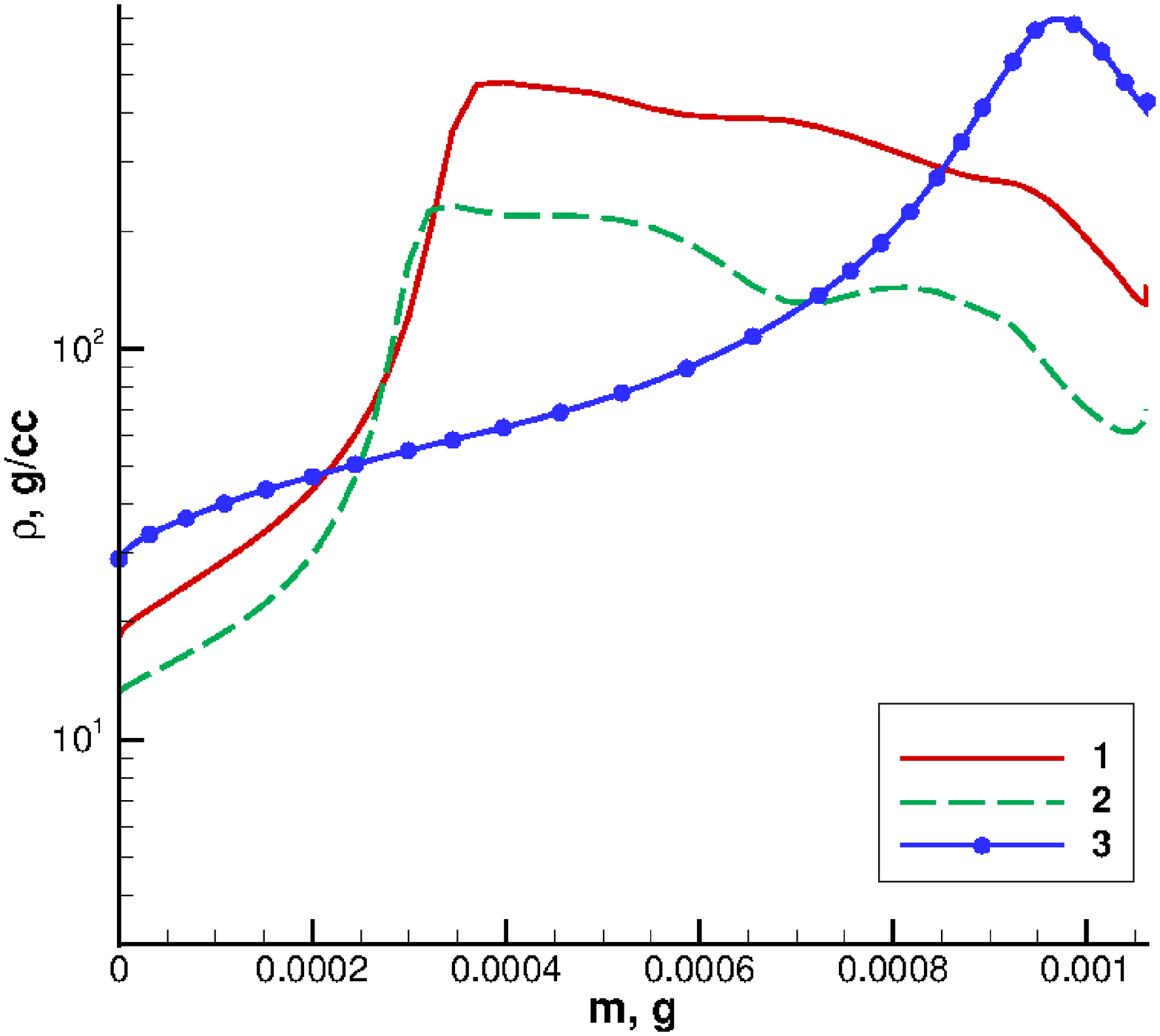}
            \put(20,84){(a)}
        \end{overpic}
    \end{minipage}
    \quad
    \begin{minipage}[b]{0.45\textwidth}
        \begin{overpic}[width=\textwidth]{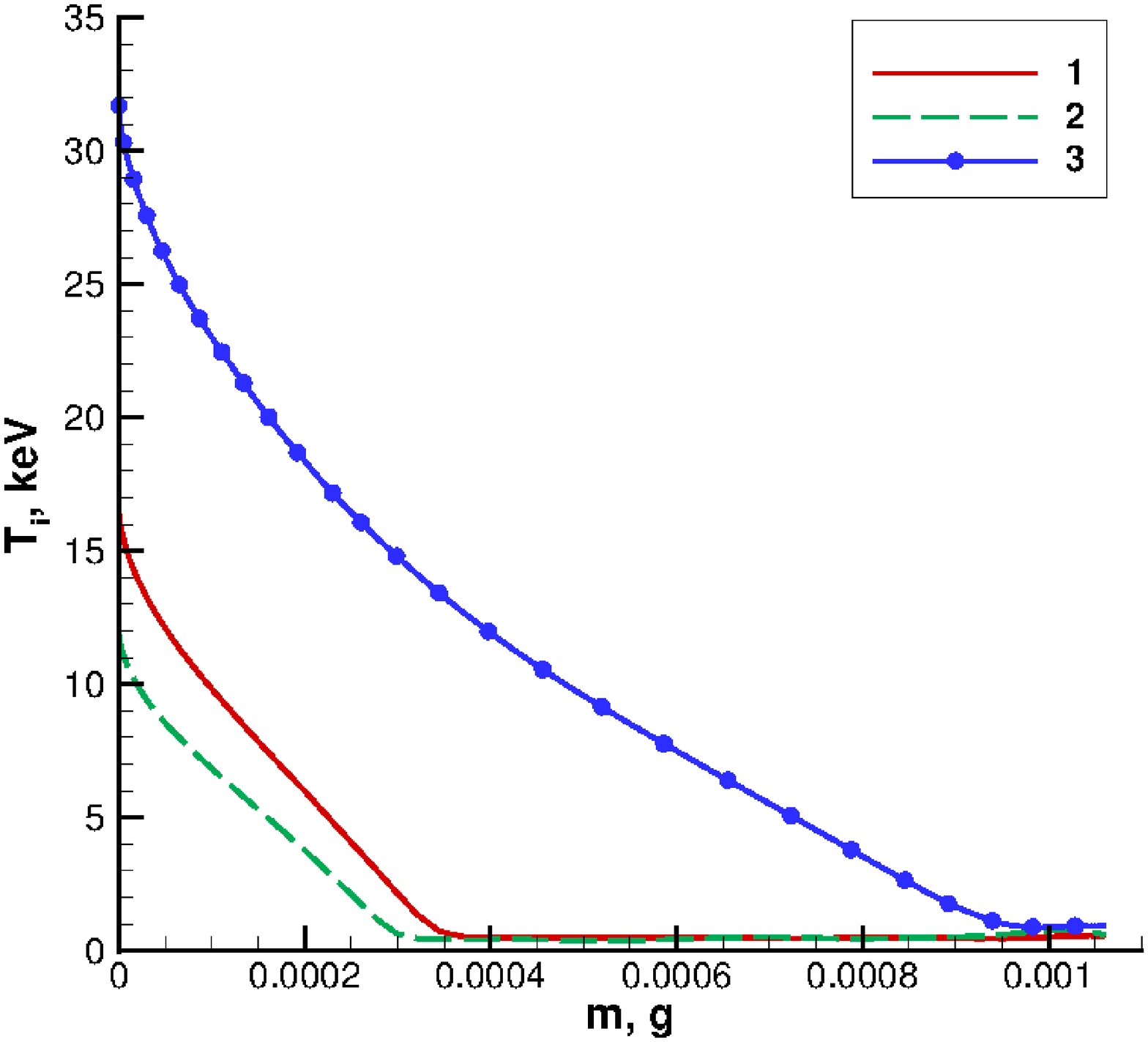}
            \put(20,84){(b)}
        \end{overpic}
    \end{minipage}
    \caption{The density (a) and temperature (b) distributions on DT-fuel mass at the time of maximum compression of
    DT-layer $t_c$ in the case of $T_h = 30$~keV and $\eta = 0.01$. Curve 1 corresponds to calculation, which takes into
    account the `wandering' effect ($t_c = 11.02$~ns, $G = 5.7$), 2 -- without the `wandering' effect ($t_c = 11.033$~ns,
     $G = 1.03$), 3 -- without fast electron generation ($t_c = 11.052$~ns, $G = 21$).}\label{fig:008}
\end{figure}

The results of numerical calculations explaining the nature of target's ignition are shown in Fig.~\ref{fig:008}, where the density and ion temperature distributions on DT-fuel mass are shown at the time moment close to onset of burning the target with and without taking into account the `wandering' effect. These data refer to calculations performed with $\eta = 0.01$ and $T_h = 30$~keV. The target ignites in both cases, but at the same time the negative effect of energy transfer by fast electrons is largely manifested. The figures also show the data calculated in the absence of fast electron  generation. In the latter, there is a significant supply of all parameters of burning plasma as compared with the ignition threshold. The ignition occurs in the hotspot with temperature of about 25~keV, density of about 40~g/cm$^3$, mass of about 0.06~mg. Combustion propagates to the surrounded dense fuel which has a mass of 20 times larger than the mass of hotspot, and where the density reaches the value 700~g/cm$^3$ that is more than 17 times larger in comparison with ignition region. In the case of `wandering' effect the conditions for ignition and burning turn out to be much worse, however, the supply relatively ignition threshold remains. In this case, the ignition occurs in a hotspot with temperature of about 12~keV, density of about 30~g/cm$^3$, and mass of about 0.1~mg. Combustion extends to the surrounded dense fuel with a mass of 10 times greater than the mass of hotspot, where the density reaches the value 500~g/cm$^3$ that  is more than 15 times larger in comparison with ignition region. The target burns worse, however, the gain is 5.7. Finally, without taking into account the `wandering' effect, there is practically no supply for the ignition conditions. The hotspot has density of about 20~g/cm$^3$, temperature of only about 4~keV and mass of about 0.25~mg. The maximum density of surrounded fuel is only 150~g/cm$^3$, and its mass is only 4 times larger than the mass of the hotspot. The target with such parameters is on the verge of ignition. The gain is 1.03. As a characteristic of ICF target implosion isentropicity, the implosion adiabat $\alpha$ is often used (see, for example,~\cite{Lindl2018} and~\cite{Goncharov2014,Regan2016}), which is the pressure $P$, averaged over the mass of one or another part of the target at a characteristic time moment, to the Fermi pressure of a completely degenerated electron gas. The degradation of the efficiency of implosion associated with the effect of fast electrons is characterized by an increase in implosion adiabat. So, this value calculated for the compressed ablator at the time $t = 8.5$~ns (the middle of the time interval of the high-intensity part of the pulse) is 1.6 in the calculation in the absence of generation of fast electrons, 2.2 in the calculation taking into account the effect of `wandering' of fast electrons and 3.4 in the calculation without the effect of `wandering', when all the fast electrons produced are able to heat the compressed part of the target.

\section{Conclusion\label{sec:004}}

The `wandering' effect due to the remoteness of the region of fast electron generation from the ablation surface of imploded direct drive target leads to decreasing the negative effect of fast electron energy transfer on thermonuclear
gain. Without taking into account the `wandering' effect ignition does not occur at the temperature of fast electrons exceeding $T_h \approx 34$~keV,
for any fraction of laser energy contained in fast electrons from the range~(\ref{eq:002}) under consideration~(\ref{eq:002})
$5 \cdot 10^{-3} \leq \eta \leq 1.5 \cdot 10^{-2}$.
Referring to the discussion of the features of fast electron generation in the second section of the paper, it can be assumed that this ignition window may relate more to the case of 3rd harmonic of the Nd-laser radiation than to the case of the 2nd harmonic radiation.
Taking into account the `wandering' effect leads to the fact that the
window of the characteristics of fast electrons corresponding to ignition is significantly expanded. However, there is the
range of $T_h$ and $\eta$ values in which the target does not ignite, even if the `wandering' effect is taken into account.
The non-ignition region can be defined as $\eta > 0.006$, $35 < T_h < 55$~keV.
This non-ignition window can refer to both the case of using the 2nd and 3rd harmonics of the Nd-laser radiation.
An important circumstance is the fact that
the `wandering' effect results in the presence of ignition window with an increase in the temperature of fast electrons
above $55-60$~keV at the all values of $\eta$ less than 0.01. This ignition window corresponds to the gain $G \approx 4-8$. This ignition window can refer to the 2nd harmonic of the Nd laser radiation.

Based on the obtained results, it is possible to draw general conclusions about the effect of fast electrons on ignition.
At a temperature of $T_h < 30$~keV, fast electrons, practically, do not influence on compression and gain of a thermonuclear
target, which is expected to ignite with a laser energy of about 2~MJ. If fast electron temperature is in the most dangerous
range of $35 < T_h < 60$~keV in terms of ignition failure, there are two ways to change the parameters of target to achieve
the ignition. The first approach is fairly obvious.
In this way, high values of gain about $G \approx 20$, close to the case of the absence of fast electron generation, can be achieved. Without changing the aspect ratio, this method is associated with an increase in laser energy. The second method is nontrivial. It is, on the contrary, to reduce the thickness of the ablator in order to move to the ignition region with a moderate gain f of $G \approx 4-8$ -- into the area to the right of the region of the first gain minimum on Fig.~\ref{fig:003}. This requires less laser energy. But to achieve these moderate gain a less
laser energy would be required also in the case without fast electron generation. The question here is how much laser energy
will be greater in terms of fast electron generation in comparison with the hypothetical case of fast electron absence.
It should be noted that for both approaches, optimization of the parameters of target and laser pulse to suppress or reduce
the negative effect of fast electrons on the gain seems to be enough delicate matter. This is due to the fact that when
the target's thickness changes and, as a consequence, its  radius and mass change too, the intensity of the laser pulse
changes that means in turn the characteristics of fast electrons will change.
In addition, for high effectiveness of an ablative stabilization~\cite{Takabe1985} of hydrodynamic instabilities in both cases it is necessary to control the dynamics of target's substance evaporation, so that during the action of laser pulse almost the entire mass of the ablator has been evaporated. With a decrease in the thickness of the ablator, this should also be accompanied by a control of the heating of the compressed part of the target by hard X-ray quanta from the corona.  The method of such a control, as well as a general way to reduce the heating of the target by fast electrons, can be the use of a thin layer of a heavy substance (with a mass much smaller than the mass of the ablator) placed in the region of the cold compressed ablator near the final position of evaporation boundary.
Finally, it should be noted that at any
fast electron temperatures above $30-35$~keV, the critical limit for the fraction of laser energy contained in fast electrons
energy is $\eta \approx 0.01$. Exceeding of this value leads to non-ignition of the target at any temperature of fast
electrons with the exception of small values less than $30-35$~keV.

\ack\label{sec:007}
This research was financially supported by Russian Science Foundation under the project No.~16-11-10174.

\section*{References}

\bibliographystyle{iopart-num}
\bibliography{refs}

\end{document}